\documentclass[11pt,conference]{IEEEtran}

\usepackage{graphicx}
\usepackage[latin1]{inputenc}
\usepackage{amsmath} 
\usepackage{amssymb} 
\usepackage{amsfonts}

\newtheorem{theorem}{Theorem}
\newtheorem{definition}{Definition}

\newcommand{\F}{\textbf{F}}
\newcommand{\Z}{\textbf{Z}}
\newcommand{\FF}{\textbf{F}_{2^m}}
\newcommand{\xx}{\textbf{x}}
\newcommand{\zero}{\textbf{0}}
\newcommand{\yy}{\textbf{y}}
\newcommand{\uu}{\textbf{u}}
\newcommand{\ee}{\textbf{e}}
\newcommand{\cc}{\textbf{c}}
\newcommand{\myalpha}{\boldsymbol{\alpha}}

\begin{document}

\title{Efficient erasure decoding of Reed-Solomon codes}

\author{
\authorblockN{Fr\'ed\'eric Didier}
\authorblockA{
EPFL, IC - IIF - ALGO, B\^atiment BC,\\
Station 14, CH - 1015 Lausanne\\
frederic.didier@epfl.ch\\
}}

\maketitle

\begin{abstract}
We present a practical algorithm to decode erasures of Reed-Solomon
codes over the $q$ elements binary field in $O(q\log_2^2 q)$ time
where the constant implied by the $O$-notation is very small.
Asymptotically fast algorithms based on fast polynomial arithmetic 
were already known, but even if their complexity is similar,
they are mostly impractical. By comparison our algorithm uses only 
a few Walsh transforms and has been easily implemented.


\end{abstract}

\section{Introduction}

A linear error-correcting code of dimension $k$ and block length $n$ over a finite field 
$\F_q$ is a $k$-dimensional linear subspace of the space $\F_q^n$. Elements of this 
subspace are called codewords. A linear code also comes with an encoding function 
that maps in a unique way an element of $\F_q^k$ (the message) into a codeword.
By erasure decoding, we mean the task of recovering the message knowing only
a subset of the coordinates of its encoding.

Of course, if we know fewer than $k$ coordinates of a codeword, there is
always more than one possible corresponding message. Thus, a code is
optimal with respect to recovering  erasures if given any subset of $k$
coordinates of a codeword, there is only one possible corresponding
message. Such a code is called maximum distance separable (MDS) code. A
standard and famous class of MDS codes is given by Reed-Solomon codes
\cite{MSbook}. Such a code is obtained by evaluating polynomials over
$\F_q$  with degree less than $k$ at $n$ different points in $\F_q$. Their
length $n$ is thus bounded by the number of points in $\F_q$, that is $q$.

Decoding erasures of any linear code can be done by a simple Gaussian
elimination with $O(k^3)$ operations. For Reed-Solomon codes, classical
algorithms can decode in $O(k^2)$ and encode in $O(kn)$. Theoretically,
using fast polynomial arithmetic \cite{AHUbook}, we can encode and decode
them in
$O(n\log^2n\log\log n)$ \cite[p. 369]{MSbook}. However, the algorithms
involved are complex and there is a large constant hidden in this asymptotic
complexity. Hence, from a practical point of view, only the quadratic time
algorithms were useful, see for instance \cite{blomer95}.

In this paper, we present a practical algorithm that can decode erasures of a
Reed-Solomon codes over $\F_q$ with $q=2^m$ in $O(q\log_2^2 q)$ time. 
It uses $O(q\log_2 q)$ memory, but we also have a version 
using $O(q)$ memory with complexity $O(q\log_2^3 q)$.
Here, the field operations are counted as $O(1)$ and the memory to store 
one field element is counted as $O(1)$ too.

This algorithm is simple (our full $C$  implementation is less than 500
lines of code) and has a very small constant in its complexity. It uses
the Walsh transform instead of the discrete Fourier transform used in
previous asymptotically fast algorithms. This is possible because we
actually never compute the coefficients of the involved polynomials, but
just manipulate their Lagrange form at the points we received.

Notice that we can use the same algorithm to encode Reed-Solomon codes in
a systematic way by choosing the first $k$ positions of a codeword and
then erasure decoding. Notice as well that the  complexity does not depend
on $k$ or $n$, so we better use Reed-Solomon codes of length close to $q$
and a dimension $k$ of the same order. In the case where only a few
systematic symbols are missing,  we can also decode in $O(q \log_2 q)$
operations plus $O(k)$ operations per missing symbol.

Erasure codes with faster encoding and decoding complexity exist
\cite{LMSS01}. Such codes are binary codes and thus cannot be MDS, that
is they require a little more than $k$ symbols to be able to recover the
message. Due to their low complexity and binary nature, these codes have many
practical applications,  but in some situations it may be better to use
the classical Reed-Solomon codes.
This is in particular the case when high rate codes are needed or when
one prefers not to waste any redundancy at the price of a slightly higher
complexity.

The organization of the paper is quite straightforward. We start by
presenting our algorithm outline before discussing in detail its two main
steps. We also recall on the way basic facts about the Walsh transform.

\section{Algorithm outline}

We start by fixing some notation. We will mainly work on the binary field
$\F_{2^m}$ with $q$ elements. We will use the notation $\oplus$ for the 
addition in this field to avoid confusion with the normal addition.
Seeing this field as an $m$ dimensional space
over $\F_2$, we represent its elements as binary vectors $\xx$ of  length
$m$.

The codewords of a Reed-Solomon code of dimension $k$ over
$\FF$ are in one-to-one correspondence with the polynomials of degree less
than $k$ over $\FF$.  Given such a polynomial $P$, we will take for its
corresponding codeword the evaluation of $P$ at all the points $\xx$ of
$\FF$, that is the image vector $(P(\xx),\xx \in \FF)$. It is of course 
possible to use smaller length, but our algorithm will recover the full
image vector anyway.

We will always
order the points $\xx=(x_1,\dots,x_m)$ of $\FF$ by lexicographical
order over the binary vector $(x_1,\dots,x_m)$. Moreover, for any function
$F$ defined  over $\FF$, we will write $[F]$ for its image vector over all
the $q$ points of $\FF$ ordered by this order.

Suppose now that we see only $k$ points (or more) of a given vector $[P]$,
we will show that we can then compute all the points of this vector in
$O(q\log_2^2 q)$ time. This operation is sufficient to both encode and decode
the code. To encode in a systematic way, we can set the first $k$ positions
of the vector $[P]$ as we want and then use the algorithm to compute the
parity symbols. To decode, we will just reconstruct the vector
from any $k$ symbols and read the encoded information at the beginning of
the vector.

We will write $R \subseteq \FF$ for a set of $k$ positions among the 
received ones. Since $P$ is a polynomial of degree less than $k$, it
is uniquely determined by its values at $k$ points of the field $\FF$.
Using the well known Lagrange interpolation formula, we have:
\begin{definition}[Lagrange Form]
Given the values of a polynomial $P$ of degree less than $k$ on a subset $R$ 
of $k$ points in $\FF$, its Lagrange form is
\begin{equation}
\label{eqn:lagrange}
P(\xx) = \bigoplus_{\uu \in R} \ 
\cc_\uu \prod_{\yy \in R, \yy \ne \uu} (\xx \oplus \yy)
\end{equation}
where the $\cc_\uu$ are the Lagrange coefficients and belong to $\FF$.
\end{definition}

Our algorithm works in two steps that we will detail in the next two
sections. The first step is to compute  the coefficients $\cc_\uu$ and it
runs in time $O(q\log_2 q)$. The second step is to evaluate the Lagrange 
form of $P$ at all the points of $\FF$ and runs in time $O(q\log_2^2 q)$.
Both steps rely heavily on Walsh transform computation.

\section{Computing Lagrange coefficients}

\begin{theorem}
Given the values at $k$ points of a polynomial of degree less than $k$
over $\FF$, we can compute its Lagrange coefficients at theses points in
$O(q \log_2 q)$ time and $O(q)$ memory.
\end{theorem}

If we apply the formula \eqref{eqn:lagrange} at a point $\xx \in R$ where
we know $P(\xx)$, we have
\begin{equation}
P(\xx) = \cc_\xx \prod_{\yy \in R, \yy \ne \xx} (\xx \oplus \yy)\ \quad \forall \xx \in R\ .
\end{equation}
The main difficulty to compute the coefficients $\cc_\xx$ is then to evaluate the product,
that is
\begin{equation}
\label{eqn:prod}
\Pi(\xx) := \prod_{\yy \in R, \yy \ne \xx} (\xx \oplus \yy) \ .
\end{equation}
We define a function $R$ from $\FF$
into $\{0,1\}$ such that $R(\xx)=1$ if and only if $\xx \in R$. We can see
this function as the indicator function of the received positions. With
this definition, we can rewrite the product \eqref{eqn:prod} as
\begin{equation}
\Pi(\xx) = \prod_{\yy \in \FF, \yy \ne \xx} (\xx \oplus \yy)^{R(\yy)}\ .
\end{equation}
By fixing a primitive element $\myalpha$ of the multiplicative group
$\FF^*$, we can also define the discrete logarithm function $L:\FF^*
\rightarrow [0,q-1]$ such that $L(\xx)=i$ if and only if $\xx=\myalpha^i$.
If we extend this function to $\FF$ by setting $L(\zero):=0$ we have
\begin{equation}
\label{eqn:log}
L(\Pi(\xx)) = \sum_{\yy \in \FF} R(\yy) L(\xx \oplus \yy)\ .
\end{equation}
and more importantly $\myalpha^{L(\Pi(\xx))}=\Pi(\xx)$ for all $\xx$ in $\FF$.
This is because $\Pi(\xx)$ is never equal to zero.
For someone familiar with the Walsh transform, it is now easy to notice that
the value of the expression \eqref{eqn:log} can be computed for all $\xx$ in
$O(q\log_2 q)$ operations.
Of course we first have to compute the vector $[L]$ but this can be done 
in linear time and we can even precompute its Walsh transform.

Since the Walsh transform plays a key role in this paper, we recall how it
works. The Walsh transform $\widehat{R}$ of a function $R:\FF \rightarrow
\Z$ is a linear transform defined by
\begin{equation}
\widehat{R}(\xx) = \sum_{\yy \in \FF} R(\yy)(-1)^{\xx \cdot \yy} \ .
\end{equation}
Moreover, we have the inverse formula
\begin{equation}
R(\xx) = \frac{1}{q} \widehat{\widehat{R}}(\xx)
\end{equation}
and for two functions $R$ and $L$ from $\FF$ into $\Z$, if we define
the convolution product $*$ by
\begin{equation}
(R*L)(\xx) := \sum_{\yy \in \FF} R(\yy)L(\xx \oplus \yy)
\end{equation}
we have
\begin{equation}
\widehat{R*L} = \widehat{R}\widehat{L} \ .
\end{equation}
So we can compute \eqref{eqn:log} by performing three Walsh transforms.
Finally, the Walsh transform can be computed efficiently in $O(q\log_2 q)$
by working on the image vector $[R]$ of a function thanks to the induction
relation
\begin{equation}
[\widehat{R}] = 
[\widehat{R_0}-\widehat{R_1} \mid \widehat{R_0}+\widehat{R_1} ]
\end{equation}
where both $R_0$ and $R_1$ are functions from $\F_{2^{m-1}}$ into $\Z$ defined
by $R_0(x_1,\dots,x_{m-1}):=R(0,x_1,\dots,x_{m-1})$ and 
$R_1(x_1,\dots,x_{m-1}):=R(1,x_1,\dots,x_{m-1})$.
This algorithm is known as the fast Walsh transform.

We remark that for computing the Lagrange coefficients, we are only
interested in the values modulo $q-1$. Since $q$ is equal to 1 modulo
$q-1$ we can perform all the above computation modulo $q-1$. With this
tweak, the Walsh transform becomes involutive, and we do not need more
than $m$ bits per value.

\section{Evaluating a Lagrange form}

\begin{theorem}
Given the Lagrange form of a polynomial $P$ over $\FF$, we can compute
its image vector $[P]$ in $O(q \log_2^2 q)$ time using $O(q\log_2 q)$ memory.
\end{theorem}

The idea to achieve this complexity start by rewriting the Lagrange
form \eqref{eqn:lagrange} of $P$ as
\begin{equation}
P(\xx) = \Pi(\xx) \bigoplus_{\yy \in R} 
\frac{\cc_\yy}{\xx\oplus\yy} \quad \forall \xx \notin R\ .
\end{equation}
To write this more conveniently, let us define an inverse function 
$I:\FF \to \FF$ that maps $\xx$ to $\xx^{-1}$ and $\zero$ to $\zero$.
We define as well a coefficient function $C$ from $\FF$ into $\FF$ that
maps $\xx$ to $\cc_\xx$ if $\xx\in R$ and to $\zero$ otherwise. We then
have
\begin{equation}
\label{eqn:big}
P(\xx) = \Pi(\xx) \bigoplus_{\yy \in \FF} C(\yy)I(\xx\oplus\yy) \ .
\end{equation}
We already computed in the previous section the vector $[\Pi]$,
so the only work left is to compute the sum on the right.
This really looks like the convolution product defined in the last
section except the functions now take their values in $\FF$ and not
in $\Z$. To overcome this difficulty, we are going to look at the
vectorial representation of the elements of $\FF$ over $\F_2$.

We will need some more notation. For $i\in\{1,\dots,m\}$, we will write 
$\ee_i$ for the $i$-th elementary basis element of the vectorial space 
$\FF$ over $\F_2$.
For a function $C$ from $\FF$ into $\FF$, we will write $C_i$ for its $i$-th
component, that is the function that maps $\xx$ into the coefficient of $\ee_i$ in
$C(\xx)$.
We can now rewrite \eqref{eqn:big} as
\begin{equation}
P(\xx) = \Pi(\xx) \bigoplus_{i=1}^m \bigoplus_{j=1}^m 
\ee_i \ee_j
\bigoplus_{\yy \in \FF}  C_i(\yy)I_j(\xx\oplus\yy) \ .
\end{equation}
Now, the sum over $\yy$ that we have to compute $m^2$ times
can be computed in $O(q \log_2 q)$ time.
This is because we now have Boolean functions.
We can thus see them as functions in $\Z$, compute
\begin{equation}
\left[\frac{1}{q}\widehat{\widehat{C_i}\widehat{I_j}}\right]
\end{equation}
using the fast Walsh transform and take the parity of each values in
the resulting vector. That is, we have
\begin{equation}
\label{eqn:decomp}
P(\xx) = \Pi(\xx) \bigoplus_{i=1}^m\bigoplus_{j=1}^m \ee_i \ee_j \ 
\text{parity}\!\left(\frac{1}{q} \widehat{\widehat{C_i}\widehat{I_j}}(\xx)\right) \ .
\end{equation}
Hence, we can evaluate the Lagrange form at all the points of $\FF$ in
$O(q\log_2^3 q)$ operations. In order to be faster, we will choose
$\ee_i=\myalpha^i$ for a primitive element $\myalpha$ of $\FF^*$. We can
then rewrite the formula
\eqref{eqn:decomp} as
\begin{equation}
P(\xx) = \Pi(\xx) \!\!\bigoplus_{s=0}^{2(m-1)}\!\! \myalpha^{s} \bigoplus_{i+j=s}
\text{parity}\!\left(\frac{1}{q} \widehat{\widehat{C_i}\widehat{I_j}}(\xx)\right) \ .
\end{equation}
Now, by using the linearity of the parity function and of the Walsh transform,
we get
\begin{equation}
P(\xx) = \Pi(\xx) \!\!\bigoplus_{s=0}^{2(m-1)}\!\! \myalpha^{s} 
\text{parity}\!\left(\frac{1}{q} 
\widehat{\ \sum_{i+j=s}\!\!\widehat{C_i}\widehat{I_j}\quad }\!\!\!\!(\xx)\right) \ .
\end{equation}
If we compute and store the $[\widehat{C_i}]$ and the $[\widehat{I_j}]$
(which require $O(q\log_2 q)$ memory and $O(q \log_2^2 q)$ operations), we
can  evaluate this formula in $O(q \log_2^2 q)$. Indeed, we have to
compute $2(m-1)$ Walsh transforms of functions which have an image vector
that can be computed in $O(q\log_2 q)$.

We remark that since we only need the parity of the result, we can
perform the Walsh transforms by keeping only the $m+1$ lower bits of all
the values. Notice as well that we can precompute the $[\widehat{I_j}]$.

If the number of erased systematic symbols is small, an alternative
decoding method is to evaluate \eqref{eqn:lagrange} directly at these
erased positions. So we still use $O(q \log_2 q)$ operations to compute
the coefficients $\cc_\uu$ but then it is only $O(k)$ operation  per
erased message symbols.

Finally, an interesting fact is that $k$ 
plays only a minor role in the implementation. We can just use all the 
available points to decode and if we received more symbols than
the degree of the polynomial sent, we will just recover the sent vector.

\section{conclusion}

We presented a practical and fast algorithm to encode and decode Reed-Solomon codes
over the erasure channel. This algorithm should be more efficient in software
than the previous algorithms as soon as the size of the field is $2^{10}$, or 
even smaller.

Moreover, it allows the use of long Reed-Solomon codes which may have some
applications. For example, our implementation can encode and decode on
$\F_{2^{16}}$ in less than a second or on $\F_{2^{20}}$ in a few seconds
(on an Intel core 2, 1.86GHz).  One may wonder where we can use such
codes, but they may be worth considering, for example, to recover from
failure on a storage system  or to send big files over the Internet.

Finally, we considered only the case of a binary field. The same approach, namely computing
Lagrange coefficients and evaluating the Lagrange form directly is certainly applicable on 
any field. However, generalizing the way we use Walsh transform is not straightforward and
it seems unlikely that it will lead to a practical algorithm.

\section*{Acknowledgment} 
The author would like to thank Amin Shokrollahi for his help to improve
this paper.

\bibliographystyle{plain}
\bibliography{RS} 

\end{document}